\documentclass[11pt]{article}
\usepackage{latexsym}
\usepackage{epsfig}

\def\bg#1{\mbox{\boldmath$#1$}}

\newcommand{\anabla}{{\overrightarrow{\nabla}}\!\!\!\!\!\!{\overleftarrow{\nabla}}}

\newcommand{\del}{\partial}
\newcommand{\beq}{\begin{eqnarray}}
\newcommand{\eeq}{\end{eqnarray}}
\newcommand{\be}{\begin{eqnarray*}}
\newcommand{\ee}{\end{eqnarray*}}
\newcommand{\bk}{{\bf k}}
\newcommand{\bp}{{\bf p}}
\newcommand{\bq}{{\bf q}}
\newcommand{\br}{{\bf r}}

\newcommand{\ra}{\rightarrow}
\newcommand{\e}{\epsilon}

\newcommand{\nn}{\nonumber}
\newcommand{\ket}[1]{\mbox{$\mid\!#1\rangle$}}
\newcommand{\bra}[1]{\mbox{$\langle#1\!\mid$}}
\newcommand{\wh}[1]{\widehat{#1}}

\begin{document}

\centerline{\Large\bf {Coulomb Effects in Low Energy Proton-Proton Scattering}}
\vskip 10mm
\centerline{Xinwei Kong$^1$ and Finn Ravndal\footnote{On leave of absence from Institute 
            of Physics, University of Oslo, N-0316 Oslo, Norway}} 
\medskip
\centerline{\it Department of Physics and Institute of Nuclear Theory,}
\centerline{\it University of Washington, Seattle, WA 98195, U.S.A}

\bigskip
\vskip 5mm
{\bf Abstract:} {\small Using a recently developed effective field theory for
the interactions of nucleons at non-relativistic energies, we calculate non-perturbatively
Coulomb corrections to proton-proton scattering. Including the dimension-eight
derivative interaction in the PDS regularization scheme, we recover a modified form of
the Blatt-Jackson relation between the scattering lengths. The effective range receives no
corrections from the Coulomb interactions to this order. Also the case of scattering in 
channels where the Coulomb force is attractive, is considered. This is of importance
for hadronic atoms.}

\bigskip

\section{Introduction}
Effective field theories are constructed to give a complete description of interacting 
particles in terms
of only the quantum fields which can be excited below a characteristic energy scale 
$\Lambda$. Degrees of freedom of higher energies are represented by an expansion
of the Lagrangian in terms of local operators of increasing dimensions. For the nucleon
system at energies below the pion mass $m_\pi$ the effective theory will thus involve 
only the nucleon field and derivatives thereof. It was first constructed by 
Weinberg\cite{SW} and has been investigated by many others\cite{others}.

In order for the effective theory to be useful it must come with well-defined counting
rules so that one knows which operators to include and which to disregard in a
calculation. Only then can one obtain a systematic and reliable expansion in $p/\Lambda$
for any scattering amplitude characterized by the momentum $p$. In the original formulation
of the effective theory for nucleons one were faced with problems in this connection 
because of the unnaturally large $S$-wave scattering lengths. This brings a smaller
energy scale into the system and the standard regularization schemes needed to handle the 
divergences from loop integrations had difficulties reproducing consistent counting
rules\cite{prob}. A solution of these problems was subsequently given by Kaplan, Savage
and Wise with the introduction of a new regularization scheme called Power Divergence 
Subtraction (PDS) and which is a generalization of the standard MS scheme based upon
dimensional regularization\cite{KSW_1}. Essentially the same method was proposed at the same 
time by Gegelia\cite{Gegelia}. Since then Mehen and Stewart\cite{Tom} have made this off-shell 
scheme (OS) more well-defined and shown that it is in fact equivalent to the PDS scheme. 

As a first application  the electromagnetic
form factor of the deuteron was calculated\cite{KSW_2}. Since then the electromagnetic
polarizability and the Compton scattering cross section for the deuteron have been
obtained within the same framework\cite{gamma}. The inelastic process of radiative neutron 
capture on protons was first considered by Savage, Scaldeferri and Wise\cite{radcap}.  
Based on the same effective
field theory also the three-nucleon system and neutron-deuteron scattering are now under 
investigation\cite{Paulo}. To leading order all these results depend only on a few known 
parameters and are similar in structure to the effective range approximation in nuclear physics. 
Even better agreement with data can be obtained in higher orders where {\it a priori} unknown 
counterterms appear. When these are determined in one process, they can be used predictively
in others. 

In the above applications of the effective field theory to low-energy nucleon interactions
there are no complications due to electromagnetic interactions between the nucleons. This is
obviously not the case for proton-proton scattering which we have set out to reconsider within
this new framework\cite{KRscatt}. At sufficiently low energy the Coulomb repulsion between 
protons becomes strong and must be included by non-perturbative methods. It is in the
same limit that the proton-proton scattering length is determined and a very careful analysis is
needed to separate the different effects. We have already succeeded in calculating the rate
for proton-proton fusion into deuterium which is dominated by the Coulomb repulsion, using this 
field-theoretic approach\cite{KRfus}. The same physics is also 
important in other hadronic scattering processes at low energies and in the nuclear bound states 
like $^3He$. Similarly, the Coulomb force will dominate in reactions between particles of 
opposite electrical charge at sufficiently low energies. Eventually it forces such systems into 
atomic bound states. 

Experimentally, proton-proton scattering was the
first hadronic process studied with the help of accelerators and one obtained early on very
accurate data\cite{BM}. Landau and Smorodinski were the first to construct a formalism in which
the Coulomb interaction could be separated from the strong interactions\cite{LS}. This was 
completed by Bethe in terms of a generalized effective range expansion of the low-energy 
phaseshift for proton-proton scattering\cite{HAB}. The more model-dependent connection between 
the measured scattering length
and the purely hadronic interaction was obtained in the phenomenological analysis of Jackson and 
Blatt\cite{JB}. Even today this is a main reference to the electromagnetic effects involved,
something which reflects the absence of a more modern and fundamental description of this 
process at low energies. The effective theory of Kaplan, Savage and Wise represents a new 
and important step in this direction. 

In the next section we present the effective theory and give a short review of the applications 
to proton-neutron and neutron-neutron scattering where Coulomb effects are absent.
For the proton-proton system these are included in Section 3. We show that a perturbative
calculation of these effects breaks down at low energies. The scattering amplitude 
depends on the non-relativistic Coulomb propagator modified by the strong interaction and
can be calculated non-perturbatively. After PDS regularization we then obtain a 
field-theoretic derivation of the Jackson-Blatt result for the scattering length to leading
order in the theory. A similar analytical result is also found using instead a momentum cutoff 
as a regulator. Hadronic scattering channels where the Coulomb force is attractive is also 
considered within the same theory. This result is of importance for the calculation of 
energy level shifts in pionic atoms and pionium\cite{KRpi}. Our leading order results have 
been confirmed by Holstein using both standard quantum mechanics and effective field theory 
with an ordinary momentum cutoff as an ultraviolet regulator combined with numerical 
integration\cite{BH}. He has also considered the more phenomenologically important case of 
several coupled channels.  

In Section 4 we consider effective range corrections by including the dimension-eight derivative
interaction to first order in perturbation theory. We are then faced 
with new and more divergent integrals involving the Coulomb propagator. By considering the
Fourier transforms of the Coulomb wavefunctions, we manage to regularize these within the same
PDS scheme as previously used. Details of this calculation are given in an appendix. The resulting 
hadronic scattering length becomes in better agreement with what one obtains in potential models. 
It also follows that the effective range itself is not affected by Coulomb interactions to this 
order in the effective field theory.

\section{Effective theory for non-relativistic nucleons}

For nucleon momenta smaller than the pion mass, we can integrate out all other fields including
the pion field. The effective Lagrangian will only involve the nucleon field $N^T = (p,n)$ and 
derivatives thereof\cite{CRS}. It must obey the symmetries we see
in strong interactions at low energies, i.e. parity, time-reversal and Galilean invariance.
Photons are coupled to satisfy local gauge invariance and we will assume here that
isospin symmetry is not broken. The Lagrangian can then be written as a series of local
operators with increasing dimensions\cite{SW}\cite{KSW_1}\cite{CRS}. In the limit where the 
energy goes to zero, the interactions of lowest dimension dominate. For this case the relevant
Lagrangian is thus
\beq  
     {\cal L}_0 = N^\dagger\left(\del_t + {\nabla^2\over 2M}\right)N
              - C_0(N^T{\bg\Pi}N)\cdot(N^T{\bg\Pi}N)^\dagger               \label{Leff}
\eeq
where $M$ is the nucleon mass. The projection operators $\Pi_i$ enforces the correct spin and 
isospin quantum numbers in the channels under investigation. More specifically, for spin-singlet 
interactions $\Pi_i = \sigma_2\tau_2\tau_i/\sqrt{8}$ while for spin-triplet interactions 
$\Pi_i = \sigma_2\sigma_i\tau_2/\sqrt{8}$. This theory is now valid below an upper energy 
$\Lambda$ which is set by the pion mass. It is also the physical cutoff when the theory is
regularized that way.

\begin{figure}[htb]
 \begin{center}
  \epsfig{figure=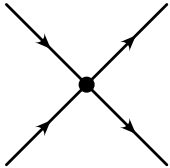,height=22mm}
 \end{center}
 \vspace{-5mm}
 \caption{\small The four-nucleon vertex corresponds to a delta-function potential.}        
 \label{fig1}
\end{figure}

The above contact interaction has dimension $D=6$ and is thus non-renormalizable. It corresponds 
to a singular delta-function potential and 
corresponds to the four-nucleon vertex in Fig.1. In order to estimate the importance
of higher order diagrams like the bubble correction to the scattering amplitude in Fig.2,
one needs counting rules. If a characteristic energy $Q$ flows through
the diagram, the propagator $1/(\omega - Q^2/2M + i\e)$ scales as $M/Q^2$ since the 
energy is typically $\omega \approx Q^2$. For the same reason the phase space factor in the
loop integration $\int d\omega d^3q/(2\pi)^4$ picks up a characteristic factor $Q^5/4\pi M$.
The estimated magnitude for the bubble diagram in Fig. 2 is thus $C_0^2 MQ/4\pi$. This will
be a perturbative correction when $C_0 MQ/4\pi < 1$ and only a few diagrams will suffice.
No real or virtual bound states will form in this case and the scattering length have
the natural size $a \approx 1/\Lambda$.

\begin{figure}[htb]
 \begin{center}
  \epsfig{figure=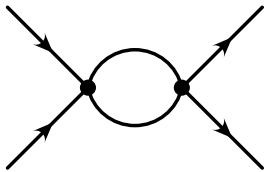,height=22mm}
 \end{center}
 \vspace{-5mm}
 \caption{\small One loop correction to the scattering amplitude.}        
 \label{fig2}
\end{figure}

More interesting is the situation when $C_0 MQ/4\pi > 1$ and the physics become 
non-perturbative. All diagrams will then contribute to the same order in $Q$ provided
the coupling constant $C_0$ runs in such a way that $C_0 \propto 1/Q$. Kaplan, Savage 
and Wise showed that a renormalization parameter $1/a < \mu \le \Lambda$ 
must then be introduced in their PDS regularization scheme\cite{KSW_1}.
The scattering length $a$ can now be unnaturally large, i.e. $a \gg 1/\Lambda$ as it is
in the nucleon-nucleon system. The full scattering amplitude $T$ due to 
the strong contact interaction is then the sum over all the chains of bubbles in Fig. 3.
\begin{figure}[htb]
 \begin{center}
  \epsfig{figure=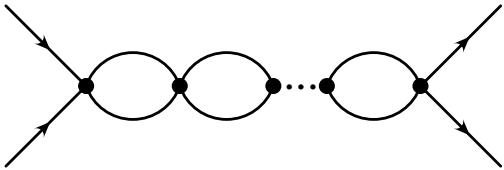,height=22mm}
 \end{center}
 \vspace{-5mm}
 \caption{\small The higher order corrections to the scattering amplitude.}        
 \label{fig3}
\end{figure}
It forms a geometric series with the sum
\beq
        T(p) = {C_0\over 1 - C_0 I_0(p)}                                  \label{scatt}
\eeq
where
\beq
     I_0(p) = \int\!{d^3q\over (2\pi)^3}{1\over E - \bq^2/M + i\e}        \label{bubble}   
\eeq
is the bubble integral and  $E = p^2/M$ is the total center-of-mass  energy. The integral is 
linearly
divergent but is finite using dimensional regularization in a space of dimension $d < 2$. It
has a simple pole when $d = 2$. In the PDS scheme one subtracts this pole which requires
the introduction of a dimensionfull parameter $\mu$. Analytically continuing back to $d=3$ 
one finds
\beq
     I_0(p,\mu) = - \left({M\over 4\pi}\right)(\mu + ip)                  \label{I0}
\eeq
The scattering amplitude (\ref{scatt}) has then the same structure has the $S$-wave partial 
wave amplitude
\beq
     T(p) = - {4\pi\over M} {1\over p\cot\delta - ip}                   \label{Swave}
\eeq
One then recovers the effective range expansion for the phase shift $\delta$ 
\beq
     p\cot\delta = - {1\over a} + {1\over 2}r_0p^2 + \ldots               \label{ERE}
\eeq
in the zero momentum limit when the coupling constant takes the renormalized value
\beq
     C_0(\mu) = {4\pi\over M} {1\over 1/a - \mu}                          \label{C0mu}
\eeq
When only the lowest order coupling parametrized by $C_0$ is included in the effective
Lagrangian (\ref{Leff}) we see that the effective range vanishes $r_0 = 0$.
The small inverse scattering length has thus a fine-tuned value given by the difference
between two large quantities. For example, in proton-neutron scattering we have 
$a_{pn} = -23.7 \,\mbox{fm}$ in the spin-singlet channel\cite{Gerry}. Choosing the value 
$\mu = m_\pi$ for the regularization parameter, we obtain $C_0 = - 3.54\,\mbox{fm}^2$. 
Physical results should be independent of the exact value of the renormalization mass $\mu$
as long as $1/a < \mu \le m_\pi$ parameter, but it strongly affects the values of the coupling 
constants whose dependence on $\mu$ is determined by the renormalization group.

For neutron-neutron collisions the scattering length is $a_{nn} = -18.4\,\mbox{fm}$, i.e. in
magnitude 25\% smaller than for the neutron-proton system. A small part of this is the
result of the proton-neutron mass difference which is a purely kinematic effect. 
The difference in the corresponding coupling
constants is however much smaller. Taking again  $\mu = m_\pi$ we see from (\ref{C0mu})
that a small relative difference $\Delta C_0/C_0$ is magnified into a larger difference
$\Delta a/a \simeq a\mu\Delta C_0/C_0$ in the scattering lengths. This is the well-known 
'amplification effect' in more standard nuclear physics where the delta-function potential 
represented by $C_0$ is replaced by a potential well of finite extension corresponding to 
$1/\mu$\cite{Ernest}. The above large difference between $a_{pn}$ and  $a_{pn}$ thus corresponds 
to a much smaller difference of less than 2\% in the coupling constants. This is a dynamical
effect due to  quark mass differences and breaking of isospin invariance by electromagnetic 
interactions at shorter scales\cite{Gerry}. Recently, Epelbaum and Meissner has shown that
these symmetry breaking effects in the nucleon scattering lengths are mostly due to the pion 
mass difference occurring in loops in higher orders of chiral perturbation theory\cite{Ulf}.

\section{Coulomb corrections to low-energy elastic scattering}

At very low energies it is the repulsive Coulomb force which dominates in proton-proton
scattering. Increasing the energy, it will still dominate in the forward and backward
directions while for other scattering angles it is overcome by the strong interaction of
very short range\cite{BM}. In our case it is described by the singular potential 
$C_0\,\delta(\br)$. The effects of transverse photons are negligible since they couple 
proportionally to the proton velocity.

\begin{figure}[htb]
 \begin{center}
  \epsfig{figure=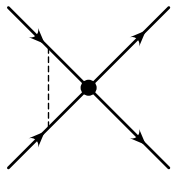,height=22mm}
 \end{center}
 \vspace{-5mm}
 \caption{\small The lowest order Coulomb correction on external legs.}        
 \label{fig4}
\end{figure}

The lowest order Coulomb correction to proton-proton scattering amplitude in Fig. 1 is 
given by the Feynman diagram in Fig. 4. Using the counting rules in the previous section, 
it is expected to give a correction $\delta T$ to the result $C_0$ from Fig.1 of the order
\beq
       \delta T \simeq C_0\left({M\over Q^2}\right)^2{e^2\over Q^2}{Q^5\over 4\pi M} 
        = C_0{\alpha M\over Q}                                     \label{esti}
\eeq
where the characteristic momentum $Q$ is here the proton momentum $p$ in the CM frame.
This should be compared with a direct calculation of the Feynman diagram which gives
\beq
      \delta T = \int\!{d^3 k\over (2\pi)^3} {e^2\over \bk^2 + \lambda^2}
                 {M\over \bp^2 - (\bp - \bk)^2 + i\e}                \label{1loop}          
\eeq
where the Coulomb photon has the mass $\lambda \ra 0$ which acts as an infrared regulator.
The integral is then finite. Combining the two denominators with the Feynman trick and using
\beq
     \int\!{d^d q\over (2\pi)^d}{1\over (\bq^2 + \Delta)^p} 
 = {\Delta^{{d\over 2} - p}\over (4\pi)^{d/2}}{\Gamma(p - d/2)\over \Gamma(p)} \label{dimint}
\eeq
we find the result
\beq
     \delta T &=&  - C_0M{ie^2\over 8\pi}\int_0^1\!dx{1\over\sqrt{x^2p^2 - (1-x)\lambda^2}} \\
             &=& -C_0 {\alpha M\over 2p}\left({\pi\over 2} + i\ln{2p\over\lambda} \right) 
              + {\cal O}(\lambda)                                   \label{T1}
\eeq
It is in agreement with the estimate in (\ref{esti}).
Since the term which depends on the photon mass is imaginary, it will not contribute to the
scattering cross section since $|C_0 + \delta T|^2 = C_0(C_0 + 2\mbox{Re}\,\delta T)$ to 
this order. The cross section is thus infrared finite and proportional to $1 - \pi\eta$
where $\eta \equiv \alpha M/2p$.
Including one more Coulomb photon exchange in the diagram Fig. 4, we find from the same
power counting rules that it will contribute a term of the order $C_0(\alpha M/Q)^2 \simeq
C_0\eta^2$. For momenta $p < \alpha M/2$ we will have $\eta > 1$ and perturbation theory
is seen to break down. The Coulomb repulsion is then strong and must be included in a 
non-perturbative way. 

\begin{figure}[htb]
 \begin{center}
  \epsfig{figure=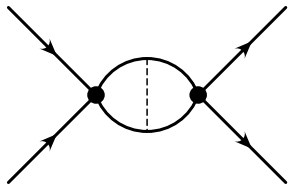,height=24mm}
 \end{center}
 \vspace{-5mm}
 \caption{\small The first Coulomb correction on the internal bubble.}        
 \label{fig5}
\end{figure}

In the same way as the external particles are strongly influenced by the repulsive
Coulomb potential at very low energies, also the interaction with the strong potential
$C_0\,\delta(\br)$ is much modified. This can already be seen in the first Coulomb
correction to the bubble diagram in Fig. 2 given by the two-loop diagram in Fig. 5. It gives 
a correction $\delta I_0$ to the integral (\ref{I0}) whose size again can be estimated
from the counting rules. Since it contains two loops and a Coulomb propagator it should be
\beq
     \delta I_0 \simeq \left({MQ\over 4\pi}\right)^2{e^2\over Q^2} = {\alpha M^2\over 4\pi}
\eeq
More accurately, it is given by the double integral
\beq
     \delta I_0 = \int\!{d^3 k\over (2\pi)^3}\int\!{d^3 q\over (2\pi)^3} 
      {M\over \bp^2 - \bq^2 + i\e} {e^2\over \bk^2 + \lambda^2}
                 {M\over \bp^2 - (\bk + \bq)^2 + i\e}                   \label{2loop}      
\eeq
It is seen to be infrared finite so that we can take the photon mass $\lambda = 0$. Since
it is logarithmic divergent in the ultraviolet, it will have a pole in $\e \equiv d - 3$ 
when it is evaluated using dimensional regularization. This is done in Appendix A where 
we find
\beq
     \delta I_0 = {\alpha M^2\over 8\pi}\left({1\over\e} + 2\ln{\mu\sqrt{\pi}\over2p} 
                  + 1 - C_E + i\pi\right)                                   \label{deltaI0}
\eeq
Here $\mu$ is the renormalization mass and $C_E = 0.5772\ldots$ is Euler's constant. The
divergent term will be removed by counterterms which modify the coupling constant
$C_0$. The prefactor which sets the magnitude of this correction, is seen to be in agreement 
with the above counting argument. In the low-energy limit where the proton momentum $p \ra 0$ 
we see again that this two-loop
correction becomes large. In order to have a finite scattering length, the logarithmic
divergent term must be cancelled by a corresponding logarithmic term coming
from the exchange of two or more photons in the bubble. These higher order Feynman diagrams
are both infrared and ultraviolet finite. Each additional photon exchange is seen from the
counting rules to bring in a factor $\alpha M/Q$ to the bubble correction. When the external
momentum goes to zero, perturbation theory breaks down again and the Coulomb-corrected
bubble must be calculated in a non-perturbative way.

\subsection{The Coulomb propagator and wavefunctions}

In relative coordinates the two-particle nucleon system is described by the free
Hamiltonian $\wh{H}_0 = \wh{\bp}^2/M$ where the reduced mass is $M/2$. The propagator
or retarded Green's function for this system is 
\beq
    \wh{G}_0^{(+)}(E) = {1\over E - \wh{H}_0 + i\e}                            \label{G0}
\eeq
where $E = p^2/M$ is the total energy. When a complete set of plane wave eigenstates 
$\ket{\bq}$ is inserted in the numerator, it becomes
\beq
  \wh{G}_0^{(+)}(E) = M\!\int\!{d^3 q\over (2\pi)^3}{\ket{\bq}\bra{\bq}\over\bp^2 -\bq^2 + i\e} 
                                                                     \label{G1}
\eeq
The propagator from an position $\br$ to a final position $\br'$ in coordinate space 
is therefore $\bra{\br'}\wh{G}_0\ket{\br} \equiv G_0(E;\br',\br)$. It follows
that the bubble diagram in Fig. 2 which is numerically given by (\ref{I0}), is just the
propagator from zero separation to zero separation, $I_0(p) = \wh{G}_0(E;0,0)$.

Including the repulsive Coulomb potential $V_C = e^2/4\pi r$ which acts between the protons,
the retarded and advanced Green's functions are now
\beq
    \wh{G}_C^{(\pm)}(E) = {1\over E - \wh{H}_0 - \wh{V}_C \pm i\e}               \label{GC}
\eeq
depending on the boundary conditions which are imposed at infinity. This is equivalent to
the integral equation $\wh{G}_C^{(\pm)} =  \wh{G}_0^{(\pm)} 
+  \wh{G}_0^{(\pm)}\wh{V}_C \wh{G}_C^{(\pm)}$. After iteration, is gives the Coulomb propagator 
as an infinite sum of Feynman diagrams with zero, one, two and so on exchanged photons as 
shown in Fig.6.

\begin{figure}[htb]
 \vspace{5mm}
 \begin{center}  
   \epsfig{figure=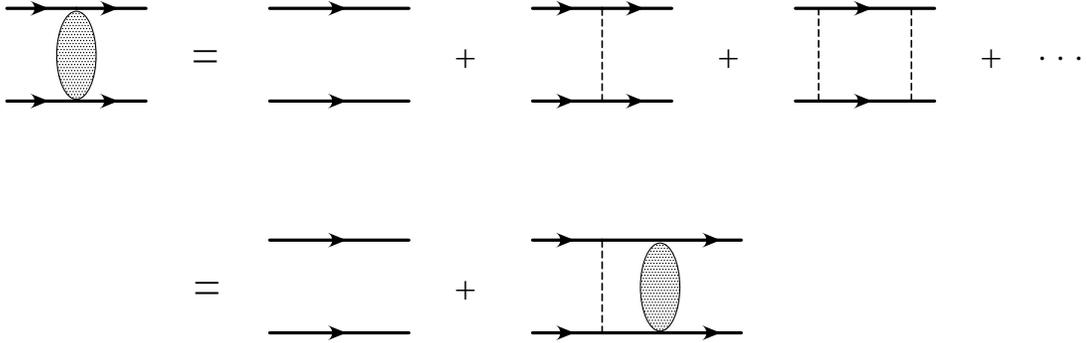,height=46mm}
 \end{center}
 \vspace{5mm}
 \caption{\small Coulomb propagator as an infinite sum.}        
 \label{fig6}
\end{figure} 

The Schr\"odinger equation 
$(\wh{H} - E)\,\ket{\psi} = 0$ where $\wh{H} = \wh{H}_0 + \wh{V}_C$ is the full Hamiltonian,
has the corresponding incoming and outgoing solutions $\ket{\psi_\bp^{(\pm)}}$. 
They can formally be expressed in terms of the free solutions as
\beq
     \ket{\psi_\bp^{(\pm)}} = [1 + \wh{G}_C^{(\pm)}\wh{V}_C]\ket{\bp}         \label{psipm}
\eeq
This is most easily seen when one uses the equivalent expression $\ket{\psi_\bp^{(\pm)}} =
\wh{G}_C^{(\pm)}\wh{G}_0^{-1}\ket{\bp}$. These solutions now have the same normalization 
as the above plane waves so that 
$\bra{\psi_{\bq}^{(\pm)}}{\mbox{$\psi_\bp^{(\pm)}\rangle$}} = (2\pi)^3\delta(\bq - \bp)$. 
Explicit solutions in coordinate space is found from solving the Schr\"odinger equation and 
can be expressed in terms of confluent hypergeometric or Kummer function 
$M(a,b;x)$\cite{LL}. For the repulsive Coulomb potential
$V_C = \alpha/r$ the in-state solution with outgoing spherical waves in the future is
\beq
     \psi_\bp^{(+)}(\br) = e^{-{1\over 2}\pi\eta}\Gamma(1 + i\eta)
                           M(-i\eta, 1; ipr - i\bp\cdot\br)\,e^{i\bp\cdot\br}    \label{psiin}
\eeq
The corresponding out-state has incoming spherical waves in the distant past and is given
by  the wavefunction 
\beq
     \psi_\bp^{(-)}(\br) = e^{-{1\over 2}\pi\eta}\Gamma(1 - i\eta)
                           M(i\eta, 1; -ipr - i\bp\cdot\br)\,e^{i\bp\cdot\br}    \label{psiout}
\eeq
where $\eta = \alpha M/2p$ is the parameter which also appeared in the earlier perturbative 
calculation. The probability to find the two protons at zero separation is thus
\beq
     C_\eta^2 \equiv
     |\psi_\bp^{(\pm)}({0})|^2 = e^{-\pi\eta}\Gamma(1 + i\eta)\,\Gamma(1 - i\eta)
                                  = {2\pi\eta\over e^{2\pi\eta} - 1}        \label{Sommer}
\eeq
which is the well-known Sommerfeld factor\cite{LL}\cite{Sommerfeld}. When the relative
velocity between the particles goes to zero, it becomes exponentially small. At higher
velocities $\eta < 1$ and the Coulomb repulsion is perturbative. We then recover to lowest 
order the result $1 - \pi\eta$ obtained from the Feynman diagram Fig. 4 in the previous 
section.

With these Coulomb eigenstates we can now find a more useful expression for
the Green's functions (\ref{GC}). Since the scattering states form a complete set in the
repulsive case we consider here, we can instead write it as in (\ref{G1}). Taking the matrix
element in coordinate space, we then have for the retarded function
\beq
     \bra{\br'}\wh{G}_C^{(+)}\ket{\br} =  M\!\int\!{d^3 q\over (2\pi)^3}
     {\psi_\bq^{(+)}(\br')\psi_\bq^{(+)*}(\br)\over\bp^2 -\bq^2 + i\e}          \label{GC1}
\eeq
In the next section we will see that this propagator gives the main part of the 
non-perturbative Coulomb corrections of the strong scattering amplitude.

\subsection{Scattering amplitudes and the modified effective range expansion}

Including the strong interaction via the local potential operator $\wh{V}_S$, 
the complete Hamiltonian becomes $\wh{H} = \wh{H}_0 +
\wh{V}_C + \wh{V}_S$. From the full Green's function $\wh{G}_{SC}^{(\pm)} = 
1/(E - \wh{H} \pm i\e)$ in the presence of both the potentials one can then formally 
construct incoming and outgoing solutions 
\beq
     \ket{\Psi_\bp^{(\pm)}} = [1 + \wh{G}_{SC}^{(\pm)}(\wh{V}_S + \wh{V}_C)]\ket{\bp} 
                                                                       \label{Psipm}
\eeq
as in (\ref{psipm}). Using now the operator identity $A^{-1} - B^{-1} = B ^{-1}(B - A)A^{-1}$
which implies the relation
\beq
     \wh{G}_{SC}^{(\pm)}-\wh{G}_{C}^{(\pm)}=\wh{G}_{C}^{(\pm)}\wh{V}_S\,\wh{G}_{SC}^{(\pm)}
\eeq
between the Green's functions, we can write the above formal solutions of the coupled
problem in terms of Coulomb states alone,
\beq
     \ket{\Psi_\bp^{(\pm)}} = [1 +
     \sum_{n=1}^\infty(\wh{G}_{C}^{(\pm)}\wh{V}_S)^n]\ket{\psi_\bp^{(\pm)}} 
                                                                       \label{Psipm1}
\eeq
The scattering amplitude is given by the $S$-matrix element which is the overlap between
an incoming state with momentum $\bp$ and an outgoing state $\bp'$. It takes the standard
form\cite{GW}
\beq
    S(\bp',\bp) &=&  \bra{\Psi_{\bp'}^{(-)}}{\mbox{$\Psi_\bp^{(+)}\rangle$}} \nn\\
                &=& (2\pi)^3\delta(\bp' - \bp) - 2\pi i\delta(E' - E) T(\bp',\bp)
\eeq
With the scattering states (\ref{Psipm1}) the $T$-matrix element can be written as the sum 
of two parts, $T(\bp',\bp) = T_C(\bp',\bp) + T_{SC}(\bp',\bp)$ where
\beq
      T_C(\bp',\bp) = \bra{\bp'}\wh{V}_C\ket{\psi_\bp^{(+)}}           \label{TC}
\eeq
is the pure Coulomb scattering amplitude and
\beq
      T_{SC}(\bp',\bp) = \bra{\psi_{\bp'}^{(-)}}\wh{V}_S\ket{\Psi_\bp^{(+)}}  \label{TSC}
\eeq
is the strong scattering amplitude modified by Coulomb corrections\cite{Harrington}.

Since the full Coulomb wavefunction $\psi_\bp^{(+)}(\br)$ is known, one can calculate exactly
the scattering amplitude (\ref{TC}). The result has the partial wave expansion\cite{LL}
\beq
     T_C(\bp',\bp) = - {4\pi\over M}\sum_{\ell = 0}^\infty\, (2\ell + 1)
                       \left[{e^{2i\sigma_\ell} - 1\over 2ip}\right]P_\ell(\cos\theta)
\eeq
where $\theta$ is the CM scattering angle and $\sigma_\ell = \mbox{arg}\Gamma(1 + \ell 
+ i\eta)$ is the Coulomb phaseshift. If the full scattering amplitude $T(\bp',\bp)$ is
defined to have the phaseshift $\sigma_\ell + \delta_\ell$, the modified strong
amplitude (\ref{TSC}) will then have the partial wave expansion
\beq
     T_{SC}(\bp',\bp) = - {4\pi\over M}\sum_{\ell = 0}^\infty \,(2\ell + 1)\,
     e^{2i\sigma_\ell}\left[{e^{2i\delta_\ell} - 1\over 2ip}\right]P_\ell(\cos\theta)
                                                                       \label{TSCp}
\eeq
It should be stressed that the phaseshift $\delta_\ell$ is not the same as one would have
in the absence of the Coulomb interaction. But it can be directly measured from the
experimental differential cross sections.

In our case the strong interaction potential $V_S = C_0\delta(\br)$ and will only affect
$S$-wave amplitude which we denote by $T_{SC}(p)$. If the corresponding phaseshift 
is called $\delta$, we see from (\ref{TSCp}) that they are related by
\beq
    p\,(\cot\delta - i) = - {4\pi\over M} {e^{2i\sigma_0}\over T_{SC}(p)} \label{Ccot}
\eeq
Because of the strong effects of the Coulomb interaction in the low-energy limit $p\ra 0$, 
the effective range expansion (\ref{ERE}) does not apply to this phaseshift. This was analyzed 
in detail by Bethe\cite{HAB} and the result can be written as the generalized expansion\cite{vanH}
\beq
    C_\eta^2 p \left(\cot\delta - i\right) + \alpha M H(\eta)
    = - {1\over a_C} + {1\over 2}r_0\,p^2 + \ldots                 \label{CERE}
\eeq
Here $C_\eta^2$ is the Sommerfeld repulsion factor (\ref{Sommer}) while  $a_C$ and
$r_0$ is respectively the $S$-wave Coulomb-modified scattering length and the effective 
range for the elastic scattering process under consideration. In the case of proton-proton 
scattering where the Coulomb potential is repulsive, the function $H(\eta)$ is
\beq
     H(\eta) = \psi(i\eta) + {1\over 2i\eta} - \ln(i\eta)                       \label{Hfun}
\eeq
where the $\psi$-function is the logarithmic derivative of the $\Gamma$-function. It
represents the effects of the Coulomb force on the strong interactions at short distances. 
Using the relation
\beq
      \mbox{Im}\psi(i\eta) = {1\over 2\eta} + {\pi\over 2}\coth\pi\eta
\eeq
we see that $\mbox{Im}H(\eta) = C_\eta^2/2\eta$ and the imaginary parts cancel out in 
(\ref{CERE}). The left-hand side will then be real and instead involve the function $h(\eta) 
= \mbox{Re}\psi(i\eta) - \ln\eta$ which is more suitable for phenomenological 
analysis\cite{JB}.

With this formalism one can extract the physical scattering length $a_C$ from  the 
experimentally measured cross-sections for proton-proton scattering at low energies\cite{JB}.
One then finds a value $a_C = -7.82$\,fm and $r_0 = 2.83$\,fm for the effective 
range\cite{Ernest}. While the effective range is essentially the same as measured in
$pn$ and $nn$ scattering, the scattering length is in magnitude less than one half the values
found in these processes. This is due to the Coulomb effects contained in $a_C$. They can
only be removed in a model for both the strong and electromagnetic interactions at short 
distances. Describing these forces in a potential model and solving the corresponding 
Schr\"odinger equation, Jackson and Blatt could isolate a strong scattering 
length $a_S$ which is determined by the measured one through their relation\cite{JB}
\beq
    {1\over a_S} =  {1\over a_C} + \alpha M
                   \left[\ln {1\over\alpha Mr_0} - 0.33\right]             \label{BJapp}
\eeq
With the above value for the effective range $r_0$ one  then finds $a_S = -17.0\,\mbox{fm}$
which is very close to the $nn$ scattering length. The term $-0.33$ in the above formula is
found to be only weakly dependent on the exact form of the strong potential\cite{Heller}.

This modified effective range expansion obviously applies also to other processes like 
$\pi^+p$ or $\pi^+\pi^+$ elastic scattering at low energies 
where repulsive Coulomb interactions are important. The only modification needed is to replace
the mass $M$ in (\ref{CERE}) with twice the reduced mass $m = m_1m_2/(m_1 + m_2)$ where
the scattered particles have different masses. Similarly, for elastic scattering in channels
like  $\pi^- p$ or $\pi^-\pi^+ $ where the Coulomb force is attractive, 
it also takes the same form. But it then involves a slightly different function
\beq
     \bar{H}(\eta) = \psi(i\eta) + {1\over 2i\eta} - \ln(-i\eta)            \label{Hbar}
\eeq
where now $\eta = - \alpha M/2p$ is negative\cite{vanH}. We will see in the following that 
both of these functions arise naturally in the present theoretical analysis.

\subsection{Coulomb corrections in the repulsive channel}

With the formalism established we can now calculated the Coulomb-modified amplitude
(\ref{TSC}) where in our case the strong interaction is represented by the contact
potential $\wh{V}_0$ with $\bra{\bp'}\wh{V}_0\ket{\bp} = C_0$. The outgoing scattering state 
$\ket{\Psi_\bp^{(+)}}$ can from (\ref{Psipm1}) be represented in terms of Coulomb states as
\beq
     \ket{\Psi_\bp^{(+)}} = \sum_{n=0}^\infty
     (\wh{G}_{C}^{(+)}\wh{V}_0)^n\ket{\psi_\bp^{(+)}}
\eeq
As a result, we then have for the scattering amplitude
\beq
    T_{SC}(p) = \sum_{n=0}^\infty\bra{\psi_\bp^{(-)}}
                \wh{V}_0\,(\wh{G}_{C}^{(+)}\wh{V}_0)^n\ket{\psi_\bp^{(+)}}  \label{Cbubbles}
\eeq
To first order in the strong coupling this is just
\beq
    T_{SC}^{(1)}(p) &=& \int\!{d^3 q\over (2\pi)^3}\int\!{d^3 q'\over (2\pi)^3}
    \bra{\psi_\bp^{(-)}}\,{\mbox{$\bq'\rangle$}}\bra{\bq'}\wh{V}_0\ket{\bq}
    {\mbox{$\langle\bq$}}\,\ket{\psi_\bp^{(+)}}\\
    &=& C_0\psi_\bp^{(-)*}(0)\psi_\bp^{(+)}(0) 
    = C_0\, C_{\eta}^2\,e^{2i\sigma_0}
\eeq
after insertions of two complete set of free states. The $S$-wave
phase shift $2\sigma_0$  comes from the relative phase between the two Coulomb 
wavefunctions (\ref{psiin}) and (\ref{psiout}). In the next order of $C_0$ we similarly 
find the contribution
\beq
     T_{SC}^{(2)}(p) =  C_0^2\, C_{\eta}^2\,e^{2i\sigma_0}J_0(p)
\eeq
where
\beq
     J_0(p) = \int\!{d^3 k\over (2\pi)^3}\int\!{d^3 k'\over (2\pi)^3}\,
              \bra{\bk'}\wh{G}_{C}^{(+)}(E)\ket{\bk} =  M\!\int\!{d^3 q\over (2\pi)^3}
     {\psi_\bq^{(+)}(0)\psi_\bq^{(+)*}(0)\over\bp^2 -\bq^2 + i\e}     \label{ICp}
\eeq
is the amplitude $\wh{G}_C^{(+)}(E;0,0)$ for the protons to propagate from initially zero 
separation and back to zero separation. It can be represented by the one bubble diagram in 
Fig.7 containing the sum of all possible exchanges of static Coulomb photons.

\begin{figure}[htb]
 \begin{center}
  \epsfig{figure=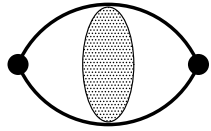,height=17mm}
 \end{center}
 \vspace{-5mm}
 \caption{\small One bubble diagram with Coulomb corrections.}        
 \label{fig7}
\end{figure}

Including higher order terms in the expansion (\ref{Cbubbles}) we see that they form a
geometric series with the sum
\beq
     T_{SC}(p) =  C_{\eta}^2{C_0\,e^{2i\sigma_0}\over 1 - C_0\,J_0(p)}    \label{TSCfin}
\eeq
This result for the scattering amplitude is now to be used in (\ref{Ccot}) which will give the
corresponding phaseshift due to the strong interaction.
The Coulomb phaseshift $\sigma_0(p)$ and the Sommerfeld factor $C_{\eta(p)}^2$ are seen to
cancel out. We are thus only left with the evaluation of (\ref{ICp}). The two 
wavefunctions gives  the Sommerfeld factor $ C_{\eta(q)}^2$ and thus we 
have
\beq
      J_0(p) = M\!\int\!{d^3 q\over (2\pi)^3} {2\pi\eta(q)\over e^{2\pi\eta(q)} - 1}
               {1\over p^2 - q^2 + i\e}                              \label{JC}
\eeq
The integral is ultraviolet divergent and must be regularized. Writing $J_0 =  J_0^{div} 
+ J_0^{fin}$ we can isolate the divergent part in
\beq
      J_0^{div} = - M\!\int\!{d^3 q\over (2\pi)^3} {2\pi\eta(q)\over e^{2\pi\eta(q)} - 1 }
       {1\over q^2}                                                       \label{ICdiv}
\eeq
which is independent of the proton momentum $p$. The remaining integral 
\beq
      J_0^{fin} = M\!\int\!{d^3 q\over (2\pi)^3} {2\pi\eta(q)\over e^{2\pi\eta(q)} - 1}
     {1\over q^2}{p^2\over p^2 - q^2 + i\e}                              \label{ICfin}
\eeq
is finite. It can be done by introducing $x = 2\pi\eta(q)$ as a new integration variable
and using
\beq
     \int_0^\infty dx {x\over (e^x - 1)(x^2 + a^2)} = {1\over 2}
     \left[\ln\left({a\over 2\pi}\right)- {\pi\over a}-\psi\left({a\over 2\pi}\right)\right]
\eeq
In our case we will have $a = 2\pi i\eta(p)$ and the finite part of the Coulomb bubble is
just the $H$-function (\ref{Hfun}) already introduced in the problem,
\beq
      J_0^{fin} = -{\alpha M^2\over 4\pi}H(\eta)                          \label{ICfinH}
\eeq
Using the full scattering amplitude (\ref{TSCfin}) in the Coulomb-modified effective range
expansion (\ref{CERE}), we see that $H(\eta)$ cancels out. The scattering length is thus 
contained in the ultraviolet divergent part of the Coulomb bubble,
\beq
      {1\over a_C} = {4\pi\over M}\left({1\over C_0} - J_0^{div}\right)   \label{appC}
\eeq
There is no contribution to the effective range $r_0$ from the Coulomb potential 
when we include only the $C_0$ interaction in the effective theory. 

We evaluate the divergent integral (\ref{ICdiv}) using dimensional regularization. With
$\e = 3 - d$ it then becomes
\beq
      J_0^{div} = - M \left({\mu\over 2}\right)^\e {\Omega_d\over (2\pi)^d}
       \int_0^\infty\!dq q^{d-3}{2\pi\eta(q)\over e^{2\pi\eta(q)} - 1}             
\eeq
where $\Omega_d = 2\pi^{d/2}/\Gamma(d/2)$ is the surface area of the $d$-dimensional unit 
sphere. Again introducing $x = 2\pi\eta(k)$ as a new integration variable, we find
\beq
      J_0^{div} &=& -M \left({\mu\over 2}\right)^\e {2\pi^{d/2}\over \Gamma(d/2)(2\pi)^d}
      (\alpha\pi M)^{d-2} \int_0^\infty\! dx {x^{\e -1}\over e^x - 1} \nn \\ 
      &=& - {\alpha M^2\over 4\sqrt{\pi}}\left({\mu\over\alpha M\sqrt{\pi}}\right)^\e
      {\Gamma(\e)\zeta(\e)\over \Gamma({3-\e\over 2})}                       \label{DR}
\eeq
Here we have introduced Riemann's zeta-function with
\beq
     \zeta(\e) = -{1\over 2}[ 1 + \e\ln 2\pi] + {\cal O}(\e^2)                 \label{zeta}
\eeq
From the factor $\Gamma(\e)$ we get a pole $1/\e$ when $\e \ra 0$. In addition there is a 
PDS pole when $d\ra 2$ from the zeta-function. Since $\zeta(\e) = \zeta(1 + 2-d) = 1/(2-d) 
+  C_E$ in this limit, it takes the form
\beq
     J_0^{div}(d\ra 2) = {\mu M\over 4\pi}{1\over d-2}                          \label{pds}
\eeq
According to the PDS regularization scheme\cite{KSW_1} this contribution should be subtracted
from the result (\ref{DR}) where then the limit $d\ra 3$ is taken. In this way we are the
left with
\beq
      J_0^{div} =  {\alpha M^2\over 4\pi}\left[{1\over\e} + \ln{\mu\sqrt{\pi}\over\alpha M}
                  + 1 - {3\over 2}C_E\right] - {\mu M\over 4\pi}         \label{JC.div}
\eeq
for the divergent part.

We can now use this result in the expression (\ref{appC}) for the measured proton-proton 
scattering length and obtain
\beq
    {1\over a_C} = {4\pi\over M C_0} + \mu - \alpha M\left[{1\over\e} + 
    \ln{\mu\sqrt{\pi}\over\alpha M} + 1 - {3\over 2}C_E\right]             \label{appC1}
\eeq
The ultraviolet pole $1/\e$ must cancel against counterterms which describe short-distance 
electromagnetic and other isospin-breaking interactions due to quark mass differences\cite{Ulf}. 
They will modify the coupling constant $C_0$ which then takes the renormalized value $C_0(\mu)$. 
It can be used to define a new scattering length
\beq
    {1\over a(\mu)} = {4\pi\over MC_0(\mu)} + \mu                         \label{app}
\eeq
It is not physical in the sense that it can be measured directly and will thus in general
depend on the renormalization point $\mu$. Coulomb effects on length scales $> 1/\mu$ have
been removed from it. The coupling constant $C_0(\mu)$ should be within a few percent
of the corresponding coupling constants for $pn$ and $nn$ scattering. From (\ref{appC1}) we 
see that $a(\mu)$ is related to the physical scattering length $a_C$ by
\beq
    {1\over a(\mu)} = {1\over a_C}  +  \alpha M
    \left[\ln{\mu\sqrt{\pi}\over\alpha M} + 1 - {3\over 2}C_E\right]             \label{Capp}
\eeq
The result is non-perturbative both in the strong coupling and in the fine structure 
constant $\alpha$ which is seen to enter in the combination $\alpha\ln\alpha$. This is a
consequence of the Coulomb force becoming strong at very low energies. Depending on the
value of the renormalization point $\mu$ we see that the Coulomb correction can actually
become of the same magnitude as the strong interaction. Since the scattering length 
$a_C$ also is negative, it can have a very big effect on the size of the hadronic scattering
length $a(\mu)$. 

Our result for the scattering length is independent of the PDS regularization scheme we have
used in the above derivation. Instead we can use a simple momentum cutoff $\Lambda$ to make
the divergent integral (\ref{ICdiv}) finite. It then becomes
\beq
     J_0^{div} = - {M\over\pi}\int_0^\Lambda\!dq {\eta(q)\over e^{2\pi\eta(q)} - 1}
\eeq
Changing integration variable to $x = 2\pi\eta$, it simplifies to
\beq
   J_0^{div} &=& -{\alpha M^2\over 2\pi}\int_{\pi\alpha M/\Lambda}^\infty 
                  {dx\over x (e^x -1)}\\ \nn
   &=& - {M\Lambda\over 2\pi^2} 
        + {\alpha M^2\over 4\pi}\left[\ln{2\Lambda\over\alpha M} - C_E\right] 
         + {\cal O}\left({\alpha M\over \Lambda}\right)
\eeq 
Again it is natural to define a strong and cutoff-dependent scattering length $a(\Lambda)$ in 
terms of a coupling constant $C_0(\Lambda)$ which absorbs the term linear in the cutoff,
\beq
     {1\over a(\Lambda)} = {4\pi\over M C_0(\Lambda)} + {2\Lambda\over\pi}
\eeq
In this regularization scheme it is now related to the physical scattering length by
\beq
     {1\over a(\Lambda)} = {1\over a_C}  +  \alpha M
                   \left[\ln{2\Lambda\over\alpha M} - C_E\right]             \label{Cap}
\eeq
which should be compared with (\ref{Capp}). This analytical result is in agreement with what 
Holstein obtained by a numerical integration\cite{BH}. Since in this effective theory the 
pions are integrated out, the magnitude of the cutoff $\Lambda$ is set by the pion mass $m_\pi$.

\subsection{Elastic scattering in the attractive channel}

We will now consider  elastic scattering of two non-relativistic particles with opposite 
electric charge. The full scattering amplitude will again be given by the infinite sum
(\ref{Cbubbles}) where now $\wh{G}_{C}^{(+)}$ is the Coulomb propagator in the attractive
channel. It will involve bound states in addition to the scattering states considered
previously in the repulsive case. Summing the infinite set of bubble diagrams we find as
in (\ref{TSCfin}) for the corresponding scattering amplitude
\beq
\bar{T}_{SC}(p) =  C_{\eta}^2{\bar{C}_0\,e^{2i\sigma_0}\over 1 - \bar{C}_0\,\bar{J}_0(p)}  
                                                                       \label{TSCanti}
\eeq
where now the Coulomb parameter $\eta = -\alpha M/2p$ is negative and $\bar{C}_0$ is the
strong coupling constant in this channel. When there are open, strong annihilation channels 
as in $p\bar{p}$, it will in general be complex. The full, Coulomb-dressed
bubble can be written as the sum  $\bar{J}_0 = \bar{J}_0^b + \bar{J}_0^s$ where the first term
\beq
    \bar{J}_0^b(p) = \sum_{n\ell} {|\psi_{n\ell}(0)|^2\over E - E_{n\ell}}
\eeq
comes from the bound states. Introducing here the scattering energy $E = p^2/M = 
\alpha^2M/4\eta^2$ and the bound state energies $E_{n\ell} = - \alpha^2M/4n^2$ together
with the probability $|\psi_{n\ell}(0)|^2 = (\alpha M)^3/(8\pi n^3)\delta_{\ell 0}$ to find 
the particles at the origin of the bound state, we obtain
\beq
    \bar{J}_0^b(p) = {\alpha M^2\over 4\pi} \sum_{n=1}^\infty {2\eta^2\over n(n^2 + \eta^2)}
             = {\alpha M^2\over 4\pi}\left[\psi(i\eta) + \psi(-i\eta) + 2C_E\right] \label{Jb}
\eeq
where again $C_E$ is Euler's constant.

The second term $\bar{J}_0^s$ in the attractive Coulomb bubble is due to the scattering 
states and has exactly the same form as (\ref{JC}) in the repulsive case. Since $\eta$ is 
now negative, we rewrite the denominator in the Sommerfeld factor as
\beq
      {1\over e^{2\pi\eta(q)} - 1} = {1\over 1 - e^{-2\pi\eta(q)}} - 1
\eeq
This factor  can then be split into three parts, $\bar{J}_0^s =
\bar{J}_0^{div} + \bar{J}_0^{fin} + \bar{J}_0^{new}$, where the two first have the same
form as in the repulsive case. In particular, we find
\beq
    \bar{J}_0^{div} = -M\!\int\!{d^3 q\over (2\pi)^3} {-2\pi\eta(q)\over e^{-2\pi\eta(q)} - 1}
       {1\over q^2}                                                      
\eeq
which is therefore given by the result (\ref{JC.div}). The $1/\e$ divergence we again
absorb into an unknown counterterm. Similarly,
\beq
     \bar{J}_0^{fin} = M\!\int\!{d^3 q\over (2\pi)^3} {-2\pi\eta(q)\over e^{-2\pi\eta(q)} - 1}
     {1\over q^2}{p^2\over p^2 - q^2 + i\e}                              
\eeq
is given by the finite result (\ref{ICfinH}). The only new integral appearing here in the
attractive channel is
\beq
    \bar{J}_0^{new} = M\!\int\!{d^3 q\over (2\pi)^3} {-2\pi\eta(q)\over p^2 - q^2 + i\e}
\eeq
It is most easily evaluated by dimensional regularization which gives
\beq
    \bar{J}_0^{new} = - {\alpha M^2\over 2\pi}\left[\ln{\mu\sqrt{\pi}\over\alpha M} 
                      + \ln(-i\eta) + 1 - {1\over 2}C_E\right]           \label{Jnew}
\eeq
It does not have any PDS poles so this is the full result. Adding together these three 
scattering state contributions, we obtain
\beq
    \bar{J}_0^s = - {\mu M\over 4\pi} - {\alpha M^2\over 4\pi}
                    \left[\ln{\mu\sqrt{\pi}\over\alpha M} + \ln(-i\eta) - {1\over 2i\eta} 
                + \psi(-i\eta) + 1 +{1\over 2}C_E\right]
\eeq
Combining this with the contribution (\ref{Jb}) from the bound states, we finally have for 
the full, attractive Coulomb bubble
\beq
    \bar{J}_0(p) = - {\mu M\over 4\pi} - {\alpha M^2\over 4\pi}
    \left[\ln{\mu\sqrt{\pi}\over\alpha M} + 1 - {3\over 2}C_E - \bar{H}(\eta)\right]
\eeq
where $\bar{H}(\eta)$ is the previously defined function (\ref{Hbar}). 

From the scattering amplitude (\ref{TSCanti}) we can now obtain the physical scattering
length $\bar{a}_C$  when combined with the corresponding effective range expansion 
(\ref{CERE}) in the attractive channel. In analogy with (\ref{app}) we can define
a strong scattering length $\bar{a}(\mu)$ in this channel directly given by the strong
coupling constant,
\beq
    {1\over \bar{a}(\mu)} = {4\pi\over M\bar{C}_0(\mu)} + \mu             \label{ant}
\eeq
It can be obtained from the measured scattering length through the relation
\beq
    {1\over \bar{a}(\mu)} = {1\over \bar{a}_C}  -  \alpha M
    \left[\ln{\mu\sqrt{\pi}\over\alpha M} + 1 - {3\over 2}C_E\right]        \label{Cant}
\eeq
Since it is not a physical quantity, it depend in general on the renormalization point
$\mu$. This result can be obtained directly from the corresponding expression
(\ref{Capp}) in the repulsive channel by letting $\alpha \ra -\alpha$ in front of the 
parenthesis. It can also be 
obtained using a momentum cutoff instead of PDS. The integral (\ref{Jnew}) will then be 
slightly different and depend logarithmically on this cutoff. The net result will be as in
(\ref{Cap}), again with the opposite sign in front of the parenthesis.

The PDS and the cutoff regularization schemes give slightly different results
for the the scattering lengths, but it will have no consequence when they are used in other
physical contexts. For instance, in the attractive channel like in $\pi^+\pi^-$ or $\pi^-p$ 
one has hadronic atoms bound by the Coulomb potential. The hydrogen-like energy level 
spectrum will then be perturbed by the strong interaction. With the hadronic contact
interactions we consider her, we will obtain a shift of the $S$-states which is easily
calculated and proportional to the scattering length in lowest order\cite{Deser}.
Since it is only $\bar{a}_C$ and not $\bar{a}(\mu)$ or $\bar{a}(\Lambda)$  which have a 
physical content and can be measured 
in scattering experiments, it must be the one which determines the level shift. This 
was first shown in a potential model calculation by Trueman\cite{Trueman}
and follows also directly using the present effective field theory. This has now also
been shown by Holstein who has extended the calculation to the more realistic situation
of several coupled channels\cite{BH}.

\section{Effective-range corrections}

Although the scattering lengths $\bar{a}(\mu)$ or $\bar{a}(\Lambda)$ are regularization
scheme dependent and cannot be directly measured, one would expect that using our results
(\ref{Capp}) and (\ref{Cap}) for them in the case of proton-proton scattering, one would 
obtained values close to the measured values of the strong scattering lengths $a_{pn}$ and
$a_{nn}$. After all, we have just obtained an ${\cal O}(\alpha)$ correction. However, from
Fig.8 where we plot $a_{pp}(\mu)$ as function of the renormalization point $\mu$, we see this
the scattering length depends strongly on this variable and actually diverges when $\mu$ is
of the order of the pion mass. The same is the case with $a_{pp}(\Lambda)$ when the cutoff
$\Lambda$ increases past $m_\pi$. This is in contrast to potential calculations represented
by the Jackson-Blatt formula (\ref{BJapp}) which permits an almost unique determination of
the Coulomb-free scattering length and essentially independent of the details of the strong 
potential. One should be able to recover this result by including higher order interactions
in the effective theory and thereby make the calculation more realistic.

\begin{figure}[htb]
 \begin{center}
  \epsfig{figure=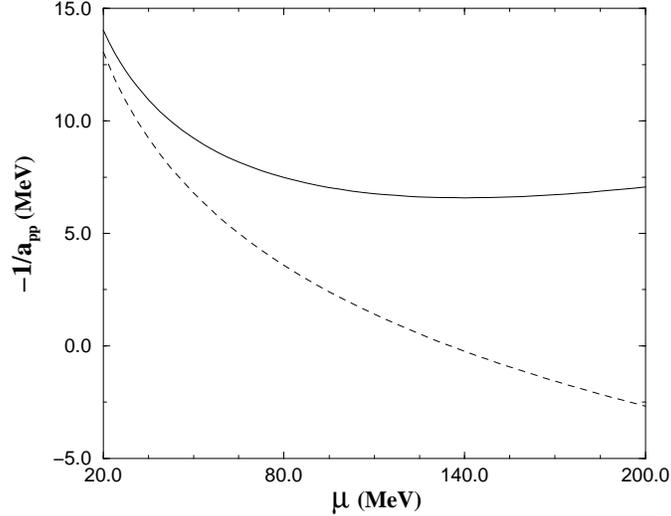,height=80mm}
 \end{center}
 \vspace{-5mm}
 \caption{\small Dependence of the inverse scattering length on the renormalization point.
                 Dashed curve gives result with only $C_0$ interaction, while the solid curve
                 also includes the $C_2$ interaction.}        
 \label{fig8}
\end{figure}

While the leading order interaction term in (\ref{Leff}) has dimension $D = 6$, the next
to leading order terms in the effective Lagrangian will have dimension $D = 8$. 
At non-relativistic energies and with only $S$-wave interactions there is only one such
term,
\beq  
     {\cal L}_2 = {1\over 2} C_2(N^T\anabla^2{\bg\Pi}N)\cdot(N^T{\bg\Pi}N)^\dagger + h.c.
                                                                     \label{L2}
\eeq
where the operator $\anabla = 1/2(\overrightarrow{\nabla} - \overleftarrow{\nabla})$.
It corresponds to a potential $\wh{V}_2$ with the matrix element
\beq
     \bra{\bq}\wh{V}_2\ket{\bk} = {C_2\over 2} (\bq^2 + \bk^2)
\eeq
Treating this operator perturbatively in the channels with no additional Coulomb 
interactions, it was found by Kaplan, Savage and Wise\cite{KSW_1} that the new coupling
constant $C_2$ is given directly in terms of the effective range $r_0$ of the scattering 
amplitude,
\beq
     C_2(\mu) = {4\pi\over M}\left({1\over 1/a - \mu}\right)^2{r_0\over 2}   \label{C2mu}
\eeq
In the previous section we saw that the leading order coupling constant $C_0$ was modified
by Coulomb effects. Here it will be shown that $C_2$ is unaffected to the order we are
working.

The initial and final scattering states are again given by the interacting states 
(\ref{Psipm1}) constructed from the pure Coulomb states (\ref{psiin}) and (\ref{psiout}).
Since we will only be concerned with $\ell = 0$ states, it is convenient to use the partial
wave expansion
\beq
   \psi_\bp^{(\pm)}(\br) = \sum_{\ell = 0}^\infty (2\ell + 1) i^\ell R_\ell^{(\pm)}(pr)
                           P_\ell(\cos\theta)               \label{partwave}
\eeq
which gives for the $S$-wave  
\beq
    R_0^{(\pm)}(pr) = C_\eta  e^{\pm i\sigma_0} e^{ipr}M(1 + i\eta,2;-2ipr)       \label{R0}
\eeq
From the Kummer identity $M(a,b;x) = e^x M(b-a,b;-x)$ it then follows that the ingoing and 
outgoing wavefunctions are simply related by complex conjugation and differ only by the phase 
factor $ e^{\pm i\sigma_0}$.

\begin{figure}[htb]
 \begin{center}
  \epsfig{figure=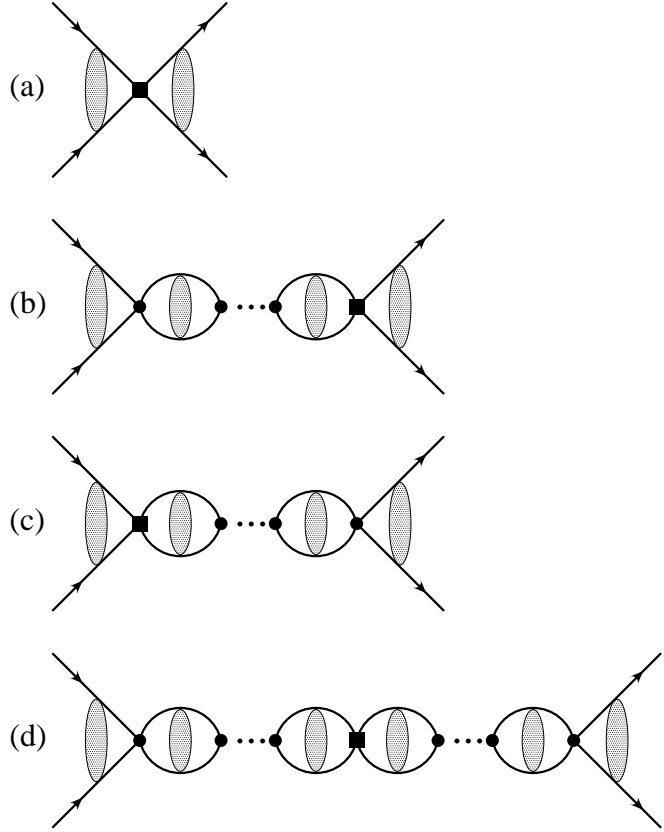,height=110mm}
 \end{center}
 \caption{\small The next to leading order diagrams.}        
 \label{fig9}
\end{figure}

Treating the new interaction (\ref{L2}) in first order of perturbation theory, we find 
the Feynman diagrams shown in Fig.9 contributing to the scattering amplitude. The first diagram 
Fig.9a gives the amount
\beq
    \delta T_{SC}^{(a)}(p) = \bra{\psi_\bp^{(-)}}\wh{V}_2\ket{\psi_\bp^{(+)}} 
     = {C_2\over 2}\!\int\!{d^3 k\over (2\pi)^3}\!\int\!{d^3 k'\over (2\pi)^3}\, 
     \psi_\bp^{(-)*}(\bk') (\bk^2 + \bk'^2)\psi_\bp^{(+)}(\bk)       
\eeq
Introducing the quantities 
\beq
    \psi_0(p) \equiv \int\!{d^3 k\over (2\pi)^3}\psi_\bp^{(+)}(\bk) 
           = C_\eta e^{i\sigma_0}                                      \label{psi_0}
\eeq
and
\beq
    \psi_2(p) \equiv \int\!{d^3 k\over (2\pi)^3}\bk^2\psi_\bp^{(+)}(\bk)   \label{psi_2}
\eeq
we thus have
\beq
     \delta T_{SC}^{(a)}(p) =  C_2 \psi_0(p)\psi_2(p)                            \label{TSCa}
\eeq
The next two chains of bubble diagrams in Fig.9b and Fig.9c form similar geometric series. 
Besides the Coulomb-dressed bubble integral (\ref{ICp}), these diagrams also involve the
related integral
\beq
     J_2(p) = \int\!{d^3 k\over (2\pi)^3}\int\!{d^3 k'\over (2\pi)^3}\,\bk^2
              \bra{\bk}\wh{G}_{C}^{(+)}(E)\ket{\bk'}                              \label{J2}
\eeq
Their sum can then be written as
\beq
     \delta T_{SC}^{(b+c)}(p) =  {C_2C_0\over 1 - C_0J_0(p)} \psi_0(p)
                                 [\psi_0(p)J_2(p) + \psi_2(p)J_0(p)]        \label{TSCbc}
\eeq
Similarly, the diagrams in Fig.9d sums up to give
\beq
     \delta T_{SC}^{(d)}(p) =  {C_2 C_0^2\over [1 - C_0J_0(p)]^2}
                                 \psi_0(p)\psi_0(p)J_0(p)J_2(p)             \label{TSCd}
\eeq
Adding up these three partial results, we then have for the perturbed scattering amplitude
\beq
     \delta T_{SC}(p) =  {C_2 \psi_0(p)\over [1 - C_0J_0(p)]^2}\left(\psi_2(p) 
   + C_0\left[\psi_0(p)J_2(p) - \psi_2(p)J_0(p)\right]\right)                \label{delTSC}
\eeq
Except for $\psi_0(p)$ which is finite and $J_0(p)$ which has already been evaluated, both
$\psi_2(p)$ and $J_2(p)$ are divergent and must be regularized. Both of them involve Coulomb
wavefunctions and there is no obvious way how to do that.

\subsection{Wavefunction regularization}

Since the coupling constants $C_0(\mu)$ and $C_2(\mu)$ in leading order are known in the
PDS regularization scheme, it would be simplest if the divergent quantities in (\ref{delTSC})
also could be regularized in the same scheme. Since this is defined in momentum space, we
will then need the Fourier transforms $\psi_\bp^{(\pm)}(\bk)$ of the Coulomb wavefunctions 
(\ref{partwave}). These were first derived by Podolsky and Pauling\cite{Linus}. Following
them, we show in Appendix B that the result can be written as
\beq
   \psi_\bp^{(\pm)}(\bk) = 4\pi \sum_{\ell = 0}^\infty (2\ell + 1) P_\ell(\cos\hat{\theta})
   \int_0^\infty\!dr r^2 R_\ell^{(\pm)}(pr)j_\ell(kr)                   \label{kwave}
\eeq
where $\hat{\theta}$ is the polar angle in momentum space. As a check, we can then verify
that $\psi_0(p)$ as defined in (\ref{psi_0}) takes its correct value. A similar calculation
in Appendix B then gives for $\psi_2(p)$ the result
\beq
    \psi_2(p) = \left[p^2 - \mu\alpha M - {1\over 2}(\alpha M)^2\right]\psi_0(p)
                                                                      \label{psi2reg}
\eeq
It is obtained using dimensional regularization and the middle, $\mu$-dependent term follows
from a PDS pole in $d=2$ dimensions.

Using the standard representation (\ref{GC1}) for the Coulomb propagator, we can now
use this result to evaluate the double integral (\ref{J2}) for $J_2$. Since the integral
over $\bk'$ equals the value of the wavefunction at the origin, we have 
\beq
     J_2(p) = M\!\int\!{d^3 k\over (2\pi)^3}\int\!{d^3 q\over (2\pi)^3}
       {\bk^2\psi_\bq^{(+)}(\bk)\psi_\bq^{(+)*}(0)\over\bp^2 -\bq^2 + i\e} 
\eeq
Now we can replace the integral over $\bk$ by $\psi_2(q)$ with the regulated result 
in (\ref{psi2reg}). This does not have to be entirely correct, replacing a part of a doubly
divergent integral with a finite, regulated expression. The problem lies in that $\psi_2(q)$ 
has in general higher order terms going to zero when $\e = 3 - d \ra 0$. However, these can 
give finite contributions when combined with the other divergence in the integral for $J_2$.
So we will proceed under the assumption that these potential terms are higher order in 
$\alpha$ so that they can be neglected within the accuracy we are working. Thus we have
\beq
J_2(p) &=& M\!\int\!{d^3 q\over (2\pi)^3} {\psi_2(q)\psi_0^*(q)\over p^2 - q^2 + i\e} \nn \\  
       &=&  [p^2 - \mu\alpha M - {1\over 2}(\alpha M)^2]J_0(p) - J   \label{J2p}
\eeq
where the last term 
\beq
     J =  M\!\int\!{d^3 q\over (2\pi)^3}{2\pi\eta(q)\over e^{2\pi\eta(q)} - 1}     \label{J}
\eeq
is independent of the external momentum $p$. As shown in Appendix B, it contains a PDS
pole  and will thus be linearly dependent on the renormalization point $\mu$. 


\subsection{Effective range and scattering length}
Since we always will choose the renormalization mass $\mu$ such that 
$\mu \gg \alpha M$, we see from (\ref{psi2reg}) that 
$\psi_2 = (p^2 - \mu \alpha M)\psi_0$. As a consequence, the last term 
$\psi_0 J_2 - \psi_2 J_0$ in (\ref{delTSC}) can be neglected compared 
to the first term. Finally we are thus left with the correction 
\beq
     \delta T_{SC}(p) =  {C_2 C_\eta^2 e^{2 i\sigma_0}\over [1 - C_0J_0(p)]^2} 
     \left[p^2 - \mu\alpha M \right]                  \label{delT}
\eeq
to the leading order scattering amplitude $T_{SC}(p)$ in (\ref{TSCfin}). The $p^2$ part
within the brackets codes for information about the effective range in proton-proton
scattering while the 
momentum-independent part will give corrections to the scattering length. In order to see
that, we need the modified effective range formula (\ref{CERE}) which now will appear in the 
form
\beq
      -{4\pi\over M} C_\eta^2 e^{2 i\sigma_0}\left({1\over T_{SC}} 
      - {\delta T_{SC}\over T_{SC}^2}\right) + \alpha MH(\eta) 
      = - {1\over a_C} + {1\over 2}r_0\,p^2 + \ldots                  \label{CEREx}
\eeq
Using now
\beq
     {\delta T_{SC}\over T_{SC}^2} = {C_2(\mu)\over C_0^2(\mu)}
     {e^{-2 i\sigma_0}\over C_\eta^2} \left[p^2 - \mu\alpha M \right]
\eeq
and the ordinary result (\ref{C2mu}) for the coupling constant $C_2(\mu)$ with the same
scattering length $a = a(\mu)$ which appears in the physical scattering length (\ref{Cant}),
we see from comparing the ${\cal O}(p^2)$ terms of this equation that the effective range
$r_0$ is not affected by the Coulomb interactions to this order in perturbation theory. This
result is also in agreement with the measured values found for $r_0$ in $nn$, $pn$ and $pp$ 
reactions\cite{Gerry}\cite{Ernest}.

However, the momentum independent terms in the correction (\ref{delT}) will modify the 
scattering length found in leading order. Instead of (\ref{Cant}) we now find
\beq
    {1\over a_{pp}(\mu)} =  {1\over a_{pp}^C} + \alpha M
    \left[\ln{\mu\sqrt{\pi}\over\alpha M} + 1 - {3\over 2}C_E 
    - {1\over 2} \mu r_0\right]                               \label{C2app}
\eeq
The new terms proportional to $r_0$ reduces the downward trend when the inverse scattering
length is plotted as function of $\mu$ as in Fig.8. When $\mu$ is around the pion mass, we
now get a value for $a(\mu)$ which is much closer to the measured values in $pn$ and $nn$
elastic scattering. In fact, we find $a(\mu = m_\pi) = - 29.9\,\mbox{fm}$ which should be 
compared with
the value  $a = -23.7 \,\mbox{fm}$ in the $pn$ channel. The differences are now within the 
range which can be explained by electromagnetic interactions at shorter scales\cite{Ulf}. 
Corrections from
higher order operators in the effective theory will have a negligible effect on this result 
since they will be associated with higher powers of $\alpha$. However, they can be slightly
different in other regularization schemes.

\section{Conclusions}

We have shown that Coulomb effects in proton-proton scattering and other hadronic systems 
at low energies can be calculated systematically in a non-perturbative way based directly
upon the full Coulomb propagator within the effective field theory of Kaplan, Savage and
Wise for nucleons. The approach is straightforward and can also be applied in other
non-relativistic field theories. The phenomenological important quantities in these systems
are the scattering lengths and effective ranges. While the Coulomb force strongly perturbs
the scattering length in proton-proton scattering, its effect in $\pi N$ and $\pi \pi$ reactions
is small due of the small reduced  mass in these systems. This is in agreement with the
experimental situation. Our results are derived in next to leading order and we find no 
changes in the effective ranges in different hadronic reactions due to the Coulomb force.
The measured values support this conclusion.

It has been shown that in higher orders in the expansion of the effective theory one is faced
with increasingly divergent integrals involving the Coulomb propagator. They have here 
been calculated
using a method based upon the Fourier transforms of the wavefunctions combined with PDS 
regularization. This approach should be put on a firmer basis or replaced by a more direct
method, perhaps in coordinate space. Also it is of interest to do these integrals 
in other regularization schemes such as the introduction of a simple momentum cutoff as we 
already have used in the simplest case.

\section{Acknowledgments} We would like to thank Martin Savage and the other members of the 
EFT group for providing an inspirational environment for this investigation. Also it is a
pleasure to thank Barry Holstein for some clarifying remarks. In addition, we are grateful to the 
Department of Physics and the INT for generous support and hospitality. Xinwei Kong is supported 
by the Research Council of Norway.

\section{Appendix A}

The two-loop integral (\ref{2loop}) which gives the lowest Coulomb correction to the
bubble diagram, is infrared finite so we can safely take $\lambda\ra 0$. It is most 
convenient to evaluate it for Euclidean external momentum $p^2 = -\gamma^2$ and write it
in $d = 3 -\e$ dimensions on the form
\beq
     \delta I_0 = \int\!{d^d k\over (2\pi)^d}\int\!{d^d q\over (2\pi)^d} 
                  {e^2 M^2\over \bq^2(\bk^2 + \gamma^2)[(\bq + \bk)^2 + \gamma^2]}
                                                             \label{twoloop}      
\eeq
after a shift of integration variables. We combine the last two factors with the Feynman
trick, integrate over $\bk$ using (\ref{dimint}) and get
\beq
    \delta I_0 = {\Gamma(2 - d/2)\over (4\pi)^{d/2}}\int_0^1\!dx\int\!{d^d q\over (2\pi)^d}
                 {e^2 M^2\over q^2\,\Delta^{2 - {d\over 2}}}
\eeq
where $\Delta = x(1-x)q^2 + \gamma^2$. Using now the more general combination formula
\beq
      {1\over A^\alpha B^\beta} = \int_0^1\!d\omega \,{\omega^{\alpha - 1}\omega^{\beta - 1}
      \over [\omega A + (1-\omega)B]^{\alpha + \beta}}
\eeq
it follows again from the general integral (\ref{dimint}) that
\beq
    \int\!{d^d q\over (2\pi)^d}{1\over q^2(q^2 + a)^{2 - d/2}} =  \int_0^1\!d\omega \,
    \omega^{1 - d/2} {\Gamma(3-d)\over \Gamma(2 - d/2)}{(\omega a)^{d-3}\over  (4\pi)^{d/2}}
\eeq
where $a = \gamma^2/x(1-x)$. Collecting the different factors, we then have for the two-loop
diagram
\beq
      \delta I_0 = e^2M^2 {\Gamma(3-d)\over  (4\pi)^d}\gamma^{2(d-3)}
                   \int_0^1\!dx\,(x-x^2)^{1 - d/2}
                   \int_0^1\!d\omega \,\omega^{{d\over 2} - 2}            \label{integrals}
\eeq
The last integral is simply $2/(1 - \e)$ while the first is
\beq
      \int_0^1\!dx\,(x-x^2)^{-{1\over 2} + {\e\over 2}}
      =   \int_0^1\!dx\,{ 1 + \e\ln\sqrt{x - x^2}\over\sqrt{x - x^2}} + {\cal O}(\e^2)
\eeq
Now
\beq
      \int_0^1\!dx\,{1\over\sqrt{x - x^2}} = \pi
\eeq
and
\beq
       \int_0^1\!dx\,{\ln\sqrt{x - x^2}\over\sqrt{x - x^2}} = - 2\pi\ln{2}
\eeq
We can then take the limit $\e \ra 0$ in (\ref{integrals}) and thus obtain
\beq
     \delta I_0 = {\alpha M^2\over 8\pi}\left({1\over\e} + 2\ln{\mu\sqrt{\pi}\over 2\gamma} 
                  + 1 - C_E\right)                                   \label{delI0}
\eeq
where $C_E = 0.5772\ldots$ is Euler's constant. Finally, we go to back to the physical 
situation by taking $\gamma = -ip$ which gives the result (\ref{deltaI0}). 

\section{Appendix B}

We will here derive the regulated results (\ref{psi2reg}) and (\ref{J2p}) from the
Fourier transformed Coulomb wavefunctions
\beq
    \psi_\bp^{(\pm)}(\bk) = \int\!d^3r\, \psi_\bp^{(\pm)}(\br)e^{-i\bk\cdot\br}
\eeq
with $ \psi_\bp^{(\pm)}(\br)$ defined by (\ref{partwave}). Attaching a hat to the spherical 
angles of $\bk$, we have
\beq
     \bk\cdot\br = kr[\cos\theta\cos\hat{\theta} 
                 + \sin\theta\sin\hat{\theta}\cos(\phi - \hat{\phi})]
\eeq
For convenience, we choose $\hat{\phi} = 0$. The integral over the azimuthal angle $\phi$ 
then gives simply a Bessel function $J_0(-kr\sin\theta\sin\hat{\theta})$. For the integral
over the polar angle $\theta$ we use the result
\beq
     \int_0^\pi\!d\theta \sin\theta P_\ell(\cos\theta)J_0(-kr\sin\theta\sin\hat{\theta})
     e^{-ikr\cos\theta\cos\hat{\theta}} = i^\ell \sqrt{2\pi\over -kr} P_\ell(\cos\hat{\theta})
     J_{\ell + {1\over 2}}(-kr)
\eeq
from Podolsky and Pauling\cite{Linus}. With $J_\ell(-z) = (-1)^\ell J_\ell(z)$ and 
introducing the spherical Bessel functions
\beq
      j_\ell(z) = \sqrt{\pi\over 2z}J_{\ell + {1\over 2}}(z)
\eeq
we thus get the result (\ref{kwave}). 

As a check, we now calculate $\psi_0(p)$ as defined in (\ref{psi_0}). Using dimensional
regularization, it will follow from
\beq
    \psi_0(p) = \left({\mu\over 2}\right)^\e \int\!{d^d k\over (2\pi)^d}\psi_\bp^{(+)}(\bk) 
                                                                \label{Psi_0}
\eeq
in the limit $\e = 3 - d \ra 0$. The angular integration will pick out the $\ell = 0$ part of 
the wavefunction and give a factor $4\pi$ when we take $d\ra 3$. Thus we are left with a 
regulated expression for the radial integral,
\beq
     \psi_0(p) = {2\over\pi}\int_0^\infty\! dr r^2 R_\ell(pr)\int_0^\infty\!dk k^{2-\e}j_0(kr)
                                                                 \label{psi0x}
\eeq
The integral over $\bk$ gives the result
\beq
      \int_0^\infty\!dk k^{{3\over 2}-\e} J_{1\over 2}(kr) = 2^{{3\over 2} -\e}
       r^{-{5\over 2} + \e} {\Gamma({3\over 2} - {\e\over 2})\over \Gamma({\e\over 2})}
\eeq
It is divergent in the limit $\e \ra 0$. The remaining integral over $r$ in (\ref{psi0x})
now involves the hypergeometric function (\ref{R0}) and is obtained from the more
general formula\cite{LL}
\beq
    \int_0^\infty\!dr  r^\nu e^{-\mu r} {_1F_1}(a,b;pr) = \Gamma(\nu + 1)\mu^{-\nu - 1}
     {_2F_1}(a,\nu + 1,b;p/\mu)
\eeq
Here it gives
\beq
    \int_0^\infty\!dr r^{-1 + \e} e^{ipr}M(1 + i\eta,2;-2ipr) = \Gamma(\e)(-ip)^{-\e}
     {_2F_1}(1 + i\eta,\e,2;2)
\eeq
The first factor is seen to cancel the divergence in the previous integral.
Combining the different terms with ${_2F_1}(1 + i\eta,0,2;2) = 1$, we finally have
\beq
     \psi_0(p) =  2 C_\eta e^{i\sigma_0}{\pi^{-1/2}} \Gamma({3/2})
\eeq
which indeed is the correct result (\ref{psi_0}).

The regularization of $\psi_2(p)$ in (\ref{psi_2}) now follows exactly along the same
lines and does not involve any new integrals. We then find
\be
     \psi_2(p) &=& 16 C_\eta e^{i\sigma_0}
                 {\Gamma({5\over 2} - {\e\over 2})\over \sqrt{\pi}2^\e}
                 {\Gamma(-2 + \e)\over \Gamma(-1 + {\e\over 2})}(-ip)^{2-\e} \cdot \\
               &&{_2F_1}(1 + i\eta,-2 +\e,2;2)\left[\left({\mu\over 2}\right)^\e 
                 (2\pi)^{2-d}{\pi^{d/2}\over\Gamma(d/2)}\right]
\ee
We have here kept the $d$-dimensional integration volume element in the last factor. The
result is now well-behaved in the limit $\e \ra 0$. Inserting  ${_2F_1}(1 + i\eta,-2,2;2) = 
1/3 - 2\eta^2/3$, we get
\beq
      \psi_2(p)|_{d\ra 3} =  C_\eta e^{i\sigma_0}\left[p^2 - {1\over 2}(\alpha M)^2\right]
\eeq
However, this time we see that the above expression also has a PDS pole when $d \ra 2$. 
Its residue involves  ${_2F_1}(1 + i\eta,-1,2;2) = -i\eta$ and it thus becomes 
\beq     
      \psi_2(p)|_{d\ra 2} =  C_\eta e^{i\sigma_0}{\alpha M\mu\over d - 2}
\eeq
It must be subtracted from the previous result with $d = 3$. We therefore finally have
\beq     
      \psi_2(p) = C_\eta e^{i\sigma_0}\left[p^2 - \alpha M\mu 
                - {1\over 2}(\alpha M)^2\right]
\eeq
which is the result we make use of in (\ref{delTSC}). In the main text we 
drop the last term since we will always choose $\mu \gg \alpha M$.

The integral (\ref{J}) is of exactly the same form as in (\ref{ICdiv}) and with dimensional
regularization takes the form
\beq
      J =  M \left({\mu\over 2}\right)^\e {\Omega_d\over (2\pi)^d}
       \int_0^\infty\!dq q^{d-1}{2\pi\eta(q)\over e^{2\pi\eta(q)} - 1}             
\eeq
Changing the integration variable to $x = 2\pi\eta(k)$ it becomes a standard integral
with the value
\beq
      J = M\left({\mu\over 2}\right)^\e\Omega_d\,(\pi\alpha M)^{3-\e}
          \Gamma(-2+\e)\,\zeta(-2+\e)
\eeq
Making now use of $\zeta(-2) = 0$, we see that it is finite when $\e \ra 0$. But again there 
is a PDS pole when $d\ra 2$. After having subtracted this part, we get
\beq
     J = - {\pi\over 4}\alpha^2M^3\left(\alpha M\zeta'(-2) + {\mu\over 12}\right)
\eeq
Under the assumption $\mu \gg \alpha M$ which we make in the main text,
this term again can be neglected.

\end{document}